\documentstyle[twoside,psfig]{memsait}

\def \SAIT #1 #2 {{\em Mem.\ Soc.\ Astron.\ It.\/} {\bf #1}, #2}
\def \MESS #1 #2 {{\em The Messenger\/} {\bf #1}, #2}
\def \ASTRNACH #1 #2 {{\em Astron. Nach.\/} {\bf #1}, #2}
\def \AAP #1 #2 {{\em Astron. Astrophys.\/} {\bf #1}, #2}
\def \AAL #1 #2 {{\em Astron. Astrophys. Lett.\/} {\bf #1}, L#2}
\def \AAR #1 #2 {{\em Astron. Astrophys. Rev.\/} {\bf #1}, #2}
\def \AAS #1 #2 {{\em Astron. Astrophys. Suppl. Ser.\/} {\bf #1}, #2}
\def \AJ #1 #2 {{\em Astron. J.\/} {\bf #1}, #2}
\def \ANNREV #1 #2 {{\em Ann. Rev. Astron. Astrophys.\/} {\bf #1}, #2}
\def \APJ #1 #2 {{\em Astrophys. J.\/} {\bf #1}, #2}
\def \APJL #1 #2 {{\em Astrophys. J. Lett.\/} {\bf #1}, L#2}
\def \APJS #1 #2 {{\em Astrophys. J. Suppl.\/} {\bf #1}, #2}
\def \APSS #1 #2 {{\em Astrophys. Space Sci.\/} {\bf #1}, #2}
\def \ASR #1 #2 {{\em Adv. Space Res.\/} {\bf #1}, #2}
\def \BAIC #1 #2 {{\em Bull. Astron. Inst. Czechosl.\/} {\bf #1}, #2}
\def \JSQRT #1 #2 {{\em J. Quant. Spectrosc. Radiat. Transfer\/} {\bf #1}, #2}
\def \MN #1 #2 {{\em Mon. Not. R. Astr. Soc.\/} {\bf #1}, #2}
\def \MEM #1 #2 {{\em Mem. R. Astr. Soc.\/} {\bf #1}, #2}
\def \PLR #1 #2 {{\em Phys. Lett. Rev.\/} {\bf #1}, #2}
\def \PASJ #1 #2 {{\em Publ. Astron. Soc. Japan\/} {\bf #1}, #2}
\def \PASP #1 #2 {{\em Publ. Astr. Soc. Pacific\/} {\bf #1}, #2}
\def \NAT #1 #2 {{\em Nature\/} {\bf #1}, #2}

\begin{opening}
\title{
ON THE ORIGIN OF THE CA-TI-CR 
ISOTOPIC ANOMALIES IN THE INCLUSION EK-1-4-1 OF THE ALLENDE-METEO\-RITE
}

\author{
   K.-L.~KRATZ$^1$, W.~B{\"O}HMER$^1$, C.~FREIBURGHAUS$^2$,
   P.~M\"OLLER$^3$, B.~PFEIFFER$^1$, 
   T.~RAUSCHER$^2$, F.-K.~THIELEMANN$^2$ 
}
\institute{$^1$Institut f{\"u}r Kernchemie, University Mainz, 
   Fritz-Stra{\ss}mann-Weg 2 \\ D-55128 Mainz, Germany  \\ 
   $^2$Departement f\"ur Physik und Astronomie, Universit\"at Basel,  
   Klingelbergstra\ss e 82, \\ CH-4056 Basel, Switzerland\\
   $^3$Theoretical Division, Los Alamos National Laboratory, Los Alamos,
   NM 87545, USA}
\date{} 
\end{opening}

\newcommand{\rb}[1]{\raisebox{1.5ex}[1.5ex]{#1}} 
\newcommand{\hw}{$T_{1/2}$}
\newcommand{\hwb}{$T_{1/2}$ }
\newcommand{\pn}{$P_{\rm n}$}
\newcommand{\pnb}{$P_{\rm n}$ }

\begin{document}

\oddpagefooter{}{}{} 
\evenpagefooter{}{}{} 
\ 
\bigskip

\begin{abstract}
   In the framework of our investigation to explain the nucleosynthetic 
origin of the correlated Ca-Ti-Cr isotopic anomalies in the Ca-Al-rich 
''FUN'' inclusion EK-1-4-1 of the Allende meteorite, the nuclear-physics 
basis in the neutron-rich $N=28$ region has been updated by including recent 
experimental data on $\beta$-decay properties and microscopic predictions 
of neutron-capture cross sections. Charged-particle and subsequent r-process 
calculations within an entropy-based approach were performed using a complete 
reaction network. It is shown that there exist two astrophysical scenarios 
within which the observed isotopic anomalies can be reproduced simultaneously; 
one at low entropies ($S\simeq10$) which confirms the earlier suggested 
SN~Ia mechanism, and another at high entropies ($S\simeq150$) which would be 
compatible with the neutrino-wind scenario of a SN~II.    
\end{abstract}

\section{Introduction}
About two decades ago, G.J. Wasserburg and his group at Caltech 
published two important papers (Lee 1978, Niederer 1980) on the identification
of correlated isotopic 
anomalies for Ca, Ti and Cr in peculiar inclusions of the Allende meteorite. 
These highly unusual isotopic compositions were attributed to fractionation 
(F) and unknown nuclear (UN) effects 
(Lee 1978, Niederer 1980) and therefore designated 
''{\bf FUN}'' anomalies. It was immediately concluded that those data might 
provide a clue to the late-stage nucleosynthetic processes which preceeded 
formation of the solar nebula. However, all astrophysical models existing 
at that time (standard r-process, generic mini r-process, explosive carbon 
burning, local proton irradiation of solar-system material, 
nuclear-equilibrium approaches) encountered severe difficulties when they 
tried to reproduce the observed anomalies of Allende, in particular those 
of the EK-1-4-1 inclusion; summarized by the -- slightly acquiescent -- 
statement of the authors: {\it ''We acknowledge our ignorance of the detailed 
astrophysical processes...''} 
(Niederer 1980 ) which produced the FUN anomalies.

As already suggested by Wasserburg et al. in 1977, 
a key to a better description of the non-standard abundances may be an
improved 
knowledge of the relevant nuclear-physics data around doubly-magic $^{48}$Ca. 
Following this line, in 1982 Sandler, Koonin and Fowler 
proposed, in their n$\beta$-process, specific nuclear-structure properties, 
i.e. low-lying (30 keV) s-wave neutron-capture resonances, in 
$^{46}$K(n,$\gamma$) and $^{49}$Ca(n,$\gamma$), respectively, to replace 
statistical Hauser-Feshbach (HF) cross sections applied in their reaction 
network. These resonances were deemed necessary to enhance the HF rates by an 
order 
of magnitude resulting in a depletion of the above progenitors of $^{46}$Ca 
and $^{49}$Ti, otherwise being considerably overproduced compared to 
observation.

\section{Experiments at CERN/ISOLDE}
At this stage, our Mainz--Lund--Garching collaboration became interested in 
the ''EK-1-4-1 story''. As nuclear chemists / physicists interested in 
astrophysics, our specific approach was to systematically investigate 
the influence of nuclear-structure properties of neutron-rich $_{16}$S to 
$_{19}$K isotopes on the nucleosynthetic production of their Ca and Ti 
$\beta$-decay daughters. From detailed spectroscopic studies of heavy K 
isotopes at CERN/ISOLDE, rather surprising results had been obtained, for 
example a sudden, strong increase of the $\beta$-delayed neutron decay 
branch ($P_{\rm n}$) from about 1$\%$ in $N=27$ $^{48}$K to 86$\%$ in $N=28$ 
$^{49}$K (Carraz 1982). In 1985, we reported the experimental identification 
of the low-lying s-wave resonance for $J^{\pi}=3/2^-$ $^{49}$Ca(n,$\gamma$)  
(Ziegert 1985b) requested by Sandler et al. (1982), 
from the {\it ''inverse reaction''} to neutron capture, i.e. the decay mode of 
$\beta$-delayed neutron emission of 472-ms ($J^{\pi}=0^-$) $^{50}$K. The 
$J^\pi=1^-$ state was, however, not found at the energy relevant for 
n$\beta$-process temperatures of $T\simeq 3-5\cdot10^8$ K, but somewhat higher 
at 155 keV. Furthermore, subsequent measurements of the partial decay widths 
of this state (Ziegert 1985a) 
have put constraints on the Breit-Wigner resonance neutron-capture rate, 
indicating that it will be roughly equal to the HF rate for the above 
n$\beta$-process temperatures, but may well be larger by up to a factor 7 
for $T\ge 1\cdot 10^9$, i.e. for typical r-process temperatures. In any case, 
our experimental data did not support the n$\beta$-explanation of the $^{49}$Ti 
abundance in EK-1-4-1 suggested by Sandler, Koonin and Fowler (1982). 

An extension of this work, using experimental data where available together 
with improved TDA and QRPA model predictions, definitely excluded the 
n$\beta$-process as a possible astrophysical scenario 
(Ziegert 1985a, Hillebrandt et al. 1986, Kratz et al. 1991) 
and favored explosive He-burning in supernovae (SN) with higher neutron 
exposures of about 5--7$\cdot10^{-5}$ mol$\cdot$cm$^{-3}\cdot$s as a likely 
site for the formation of the correlated Ca-Ti-Cr isotopic anomalies in 
EK-1-4-1. As a consequence of the new nucleosynthesis scenario, other -- 
more exotic -- progenitor isotopes became the ''key'' nuclides to produce 
only {\it little} $^{46}$Ca and {\it much} $^{48}$Ca at the same time, in 
order to cause the abundance ratio of $^{48}$Ca/$^{46}{\rm Ca}\simeq 250$ 
observed in EK-1-4-1. It was shown that in particular the -- at that time 
still unknown -- $\beta$-decay half-life of the neutron-magic $N=28$ 
{\it ''turning-point''} nucleus $^{44}$S (where the neutron-capture 
half-life becomes comparable to the $\beta$-decay half-life; 
$T_{1/2}({\rm n})\simeq T_{1/2}(\beta$)) would be decisive in defining 
the neutron-capture process duration, and consequently is determining the 
final amount of stable $^{46}$Ca (Thielemann et al. 1990).
With shell-model predictions for the decay of {\it spherical} $^{44}$S 
between 1.1~s (QRPA) and 0.3~s (TDA), the necessary time scale for the 
neutron irradiation was of the order of 100~ms, and the required abundance 
of the ($\alpha$,n) neutron source was 
$Y_{\alpha,{\rm n}} \simeq 2\cdot 10^{-3}$ mol$\cdot$cm$^{-3}$. It was well 
recognized that this might be hard to attain in He-zones of massive stars 
during a SN explosion. Already in 1979, Thielemann et al. (1979)
found that during explosive processing of a 3$\alpha$ layer of standard 
preshock composition, maximum abundances of $^{22}$Ne (as ($\alpha$,n) 
neutron source) of $Y_{\rm Ne}\simeq 10^{-3}$ and resulting free 
neutrons of $Y_{\rm n}\simeq 10^{-9}$ can be realized for the above requested 
process duration. This was also verified in recalculations with updated 
reaction rates in 1990 (Thielemann et al. 1990). 
Due to the strong influence of the neutron poison $^{25}$Mg produced by 
$^{22}$Ne($\alpha$,n), rather the alternative $^{18}$O and in particular 
$^{13}$C neutron sources were found to provide the best conditions; 
however, still at the limit to fulfill the above constraints. Nevertheless, 
the explosive He-burning scenario was not excluded immediately, because  
at that time considerable uncertainties still existed in the theoretical 
nuclear-data input of far-unstable progenitor isotopes. Unfortunately -- or 
''fortunately'' when considering the later, very surprising experimental 
results presented below --  we never published this status of the ''EK-1-4-1 
story'' in a refereed journal... 

\section{Experiments at GANIL/LISE}
In order to eliminate at least some of the remaining uncertainties in the 
nuclear-structure properties of potential ''turning-point'' nuclei south 
of $^{48}$Ca, in a Mainz--Orsay--Dubna--GANIL collaboration we have 
performed two experiments at the doubly-achromatic spectrometer LISE of 
GANIL. As production mechanism, projectile fragmentation of a 60~MeV/u 
$^{48}$Ca beam was used (Kratz et al. 1991, Sorlin et al. 1993, Kratz et al. 
1995, B\"ohmer 1996).
In these experiments, the gross $\beta$-decay properties $T_{1/2}$ and $P_{\rm n}$ 
of the very neutron-rich isotopes $^{43}$P, $^{42,44,45}$S, $^{44-46}$Cl,
and $^{47}$Ar could be measured for the first time (see Table I). Very 
surprising results were obtained. Compared to the model predictions used in 
our earlier astrophysical calculations, 
which -- quite naturally -- assumed {\it spherical} shapes for these 
$N\simeq28$ near-magic nuclides, the experimental $T_{1/2}$ for most of 
them (except $^{44,45}$Cl) were considerably shorter, by about a factor 3 
for $^{43}$P, $^{42}$S and $^{47}$Ar, a factor 4 for $^{45}$Cl, up to a 
factor 10 for $^{44,45}$S. Similarly, most of the measured $P_{\rm n}$ values 
were unexpected when assuming sphericity. We therefore suggested that our 
surprising results indicated an erosion of the $N=28$ shell-gap below 
$^{48}$Ca resulting in an onset of strong quadrupole deformation (Sorlin 
1993, Kratz et al. 1995). Although initially not predicted by any model of 
whatever sophistication and therefore not accepted in the nuclear-structure 
community, our ''unconventional'' ideas were rather soon confirmed by both 
theory and COULEX experiments (see, e.g. Werner 1994, Scheit 1996, 
Glasmacher 1998).

\section{Ground-State Shapes from a QRPA Parameter Study}   
The main motivation for our shell-model parameter study within the 
quasi-particle random-phase approximation (QRPA) with Folded-Yukawa 
single-particle (SP) wave functions (M\"oller and Randrup 1990) was to 
determine the nuclear shape -- spherical or deformed -- of the above 
$N\simeq 28$ isotopes exclusively from a comparison of their experimental 
and calculated gross $\beta$-decay properties $T_{1/2}$ and $P_{\rm n}$ 
(B\"ohmer 1996). 
These quantities were calculated as functions of quadrupole deformation, 
where the $\epsilon_2$-parameter was varied in small steps between 
-0.35 and +0.35, i.e. between strongly oblate and strongly prolate.

\begin{table}
\caption{
QRPA-calculation of \hwb and \pnb with  
optimized ground-state quadrupole deformation of the investigated nuclei. 
In column 2, the deformation parameter ${\varepsilon}_2$ is given which 
leads to the best agreement between the calculated \hw and \pn values 
(columns 3 and 4) and the experimental results (columns 5 and 6). 
}
 \label{optimal} 
\begin{center}
\begin{tabular*}{\textwidth}{@{}@{\extracolsep{\fill}}cccccc} \hline
          &                &
\multicolumn{2}{c}{Theoretical values}  &
\multicolumn{2}{c}{Experimental values} \\ \cline{3-4}
\cline{5-6} 
\rb{Isotope} & \rb{Optimized ${\varepsilon}_2$} &  {$T_{1/2}$[ms]} &
{$P_{\rm n}$[\%]}  &  {$T_{1/2}$[ms]} & {$P_{\rm n}$[\%]} \\  \hline
$^{43}$P  & $-0,175$ & \phantom{10}70 & 100,0   & 36$\pm$4  &
 100$\pm$10  \\ 
$^{42}$S  & $-0,300$ & 1300 & \phantom{10}0,9  & 579$\pm$46  &
    $<$4    \\  
        & $+0,300$ & 1570 & \phantom{10}0,1 \\$^{44}$S  &
$+0,325$ & \phantom{1}133 & \phantom{0}15,3   & 123$\pm$10 &
 18$\pm$3 \\ 
$^{45}$S  & $-0,200$ & \phantom{10}83 & \phantom{1}39,8   &
 72$\pm$5  &  54$\pm$6  \\  
          & $+0,125$ & \phantom{10}90 & \phantom{1}36,7 \\
$^{44}$Cl  & $-0,250$ & \phantom{1}434 & \phantom{10}1,8   &
438$\pm$19 &   $<$8 \\ 
$^{45}$Cl  & $-0,200$ & \phantom{1}392 & \phantom{0}10,4   &
412$\pm$39 &   38$\pm$5 \\ 
$^{46}$Cl  & $+0,100$ & \phantom{1}204 & \phantom{1}57,9   &
223$\pm$37  &  60$\pm$9 \\ 
$^{47}$Ar  & $-0,180$ & \phantom{1}705 & \phantom{10}0,1   &
613$\pm$197 &   $<$1 \\ 
           & $+0,125$ & \phantom{1}709 & \phantom{10}0,1 \\
\hline 
\end{tabular*} 
\end{center}
\end{table}

The nuclei considered in this study, have 15 to 18 protons and 26 to 
30 neutrons. In this low-mass region the single-proton and single-neutron 
levels are well separated by 1--5 MeV (see, e.g. the Folded-Yukawa 
SP-diagrams in M\"oller 1997). 
For both protons and neutrons, pronounced 
spherical shell-gaps exist for particle numbers 14, 20 and 28, and 
deformed gaps develop around $\epsilon_2\simeq +0.2$ for particle numbers 
16, 26 and 30 and around $\epsilon_2\simeq -0.25$ for particle number 18, 
replacing the above spherical shell-gaps. In any case, more or less 
independent of the nuclear shape, the low SP-level densities together with 
the $\beta$-decay selection rules yield only a few Gamow-Teller (GT) 
transitions. 
In fact, in the $A\simeq 45$ mass region considered here, the quantities 
\hwb and \pnb are to a large extent determined by one or two strong 
GT-transitions in the respective energy intervals, whereas the effects of 
forbidden transitions can be neglected. Thus, the expected correlation 
between the gross $\beta$-decay properties and the nuclear shape should 
be unambiguous, at least to distinguish between sphericity and collectivity. 
 
In the following, let us exemplarily discuss the results and their relation 
to our astrophysical ''turning-point'' nucleus $^{44}$S. According to the 
prediction of most global models, the neutron-magic nucleus $^{44}$S should 
be spherical in its ground state. A QRPA calculation in the spherical limit 
leads to a rather long half-life of $T_{1/2}=1.12$s and a maximum possible 
neutron emission probability of $P_{\rm n}=100$\% (Sorlin 1993, Kratz et 
al. 1995). In this case, the two (lowest-energy) 
$\nu{\rm f}_{7/2}\rightarrow\pi{\rm f}_{7/2}$ 
GT-transitions determining both gross $\beta$-decay properties lie above 
the neutron separation energy, at about 5 MeV in the daughter nucleus 
$^{44}$Cl. Much better agreement with the experimental values of 
$T_{1/2}=(123 \pm 10$) ms and $P_{\rm n}=(18\pm3$) \% are obtained if we 
perform the QRPA calculation with a prolate deformation of 
${\epsilon}_2 \simeq 0.3$. 
For this shape, GT-transitions between deformed Nilsson-orbitals of f$_{7/2}$ 
shell-model origin are shifted down in energy to near the ground state of 
$^{44}$Cl, resulting in a shorter half-life of \hw=133 ms together with a 
lower $\beta$-delayed-neutron emission probability of $P_{\rm n}=15.3$\%. 
From this comparison, one clearly 
can exclude the spherical shape for $^{44}$S. As mentioned above, in the 
meantime our conclusion has been confirmed by intermediate-energy Coulomb 
excitation experiments of B(E2)-values (Scheit 1996, Glasmacher 1998).

In the same way, the results on the non-spherical ground-state shapes 
for the other $N \simeq 28$ nuclei were obtained. Table~I summarizes the 
deformation parameters for which our QRPA calculations yielded the best 
agreement with the experimental values. In a number of cases, e.g. for  
$^{42,45}$S and $^{47}$Ar, our approach is not able to determine whether 
these isotopes are prolate or oblate deformed, but can again exclude a 
spherical shape. This general distinction between collectivity and 
sphericity is, however, of major importance as input for improved 
predictions of neutron-capture cross sections (Rauscher et al. 1998) 
entering our astrophysical network calculations discussed below.

\section{Theoretical Neutron-Capture Rates}  

Apart from nuclear masses and $\beta$-decay properties, also neutron-capture 
cross sections (${\sigma}_{\rm n}$) are important input data for calculations 
of non-equilibrium r-process-like nucleosynthesis scenarios. For typical 
astrophysical neutron energies, two reaction mechanisms have to be considered
(Rauscher 1995): (i) Capture reactions where a {\it compound nucleus} (CN) 
is created; in this case ${\sigma}_{\rm n}$(CN) values are commonly calculated  
with the statistical Hauser-Feshbach model. (ii) Reactions where a neutron 
is directly captured into bound states, i.e. {\it direct capture, (DC\/}). 
The DC-mechanism can dominate the neutron-capture for nuclei with low  
level density in the vicinity of the neutron separation energy $S_{\rm n}$. 
This is especially the case for light-mass nuclei, for isotopes close to magic 
neutron numbers, and for nuclei far from $\beta$-stability with low $S_{\rm n}$ 
values. Therefore, in the present study, in addition to CN cross sections 
(Rauscher et al. 1998) also the DC contributions of all even-even (ee) 
nuclei (because of level-density arguments one can assume to first order 
that DC will be more important for capture on ee targets than on odd-nucleon
ones) between $^{40}$S to $^{74}$Fe were calculated with the code TEDCA, as 
follows.

The theoretical cross section 
$\sigma^{th}$ is given by a sum over each final state 
$i$~(Kim et al. 1987)
\begin{equation}
\label{dc}
\sigma^{th}=\sum_i C_i^2 S_i \sigma_i^{DC} \quad.
\end{equation}
In our case the isospin Clebsch-Gordan coefficients $C_i$ are equal to 
unity. The 
spectroscopic
factors $S_i$ describe the overlap between the antisymmetrized wave 
functions of
target+n and the final state. In the case of one-nucleon capture on 
even-even deformed nuclei, the 
spectroscopic factor for capture into a state $i$, which has an occupation 
probability $v_i^2$ in the target, can be reduced to~
(Glendenning 1983)
\begin{equation}
S_i=1-v_i^2\quad.
\end{equation}
The corresponding probabilities $v_i^2$ are found by solving the 
Lipkin-Nogami pairing equations~
(M\"oller 1995).

The factors $\sigma_i^{DC}$ in Eq.~\ref{dc} are essentially 
determined by the overlap of the scattering wave function in the 
entrance channel, the bound-state wave function and the 
multipole-transition operator. The potentials needed for the calculation 
of the before-mentioned wave functions are obtained by applying the 
folding procedure. In this approach, the nuclear density of the target 
$\rho_{T}$ is folded with an energy and density dependent 
effective nucleon-nucleon interaction $w_{eff}$~
(Kobos et al. 1984)
\begin{equation}
V(E,R)=\lambda V_{F}(E,R)=\lambda \int \rho_{ T}
(\vec{r})w_{eff}(E,\rho_{T},\vert \vec{R}-\vec{r} 
\vert)\,d\vec{r}\quad,
\end{equation}
with $\vec{R}$ being the separation of the centers of mass of the two 
colliding nuclei. The interaction $w_{eff}$ is only weakly 
energy dependent in the energy range of interest~
(Oberhummer and Staudt 1991).
The density distributions $\rho_{T}$ were 
calculated from the folded-Yukawa wave functions.

The only remaining parameter $\lambda$ was determined by employing a 
parametrization of the volume integral $I$
\begin{equation}
I(E)=\frac{4\pi}{A}\int V_{F}(R,E)R^2\,dR\quad,
\end{equation}
expressed in units of MeV\,fm$^3$, and with the mass number $A$ of the 
target nucleus. Recently, the averaged volume integral $I_0$ 
was fitted to a function of mass number $A$, charge $Z$ and 
neutron number $N$ for a set of specially selected nuclei:~
(Balogh 1994, Krausmann et al. 1996)
\begin{equation}
I_0=255.13+984.85A^{-1/3}+9.52\cdot 10^6 \frac{N-Z}{A^3}\quad.
\end{equation}
Thus, the strength factor $\lambda$ can easily be computed for each 
nucleus by using
\begin{equation}
\lambda=\frac{I_0}{I}\quad.
\end{equation}
For the bound states, the parameters $\lambda$ are fixed by the requirement 
of a correct 
reproduction of the separation energies.


The calculated CN and DC cross sections as well as the deformation 
parameters used for the calculation of the single-particle levels are shown 
in Table~II. Experimentally known neutron separation energies in the 
final nuclei were taken from a recent data compilation (Audi and Wapstra 
1993); otherwise predictions from the finite-range droplet mass model FRDM 
(M\"oller 1995) were used. Furthermore, in this table $\beta$-decay half-lives 
($T_{1/2}$($\beta$)) are compared to theoretical lifetimes for neutron 
capture ($T_{1/2}$(n)). Experimental $T_{1/2}$($\beta$) were again 
taken from recent compilations (Audi and Wapstra 1993, Pfeiffer et al. 2000), 
and -- where necessary -- theoretical values were obtained by using the QRPA 
code of M\"oller and Randrup (1990). The $T_{1/2}$(n) were computed from our 
present DC+CN results with neutron number densities of $3\cdot10^{19}$ 
cm$^{-3}$ for S and Ar target isotopes and $6\cdot10^{20}$ cm$^{-3}$ for 
Ti, Cr and Fe isotopes, respectively. For the Ca nuclides, the values were
calculated earlier by Krausmann et al. (1996) with the same model.

\begin{table}
\label{dccs}
\caption{Calculated 30-keV (c.m.) Maxwellian averaged neutron-capture 
cross sections $<\sigma>_{30\,keV}$ for CN and DC. The column 
labeled ''\%'' gives the fraction of direct capture in the total cross 
section. Also shown are the deformation parameters $\epsilon_2$ and the 
neutron separation energies S$_{n}$ of the final nuclei, target+n. 
The neutron-capture half-lives $T_{1/2}$(n) were computed with the values 
from column ''DC+CN'' and neutron number densities of $3\cdot10^{19}$ for 
S and Ar target isototpes, and $6\cdot10^{20}$ cm$^{-3}$ for Ti, Cr and Fe 
isotopes, respectively. The values for the Ca isotopes were taken from an 
earlier calculation using the same model (Krausmann et al. 1996).}
\begin{center}
\begin{tabular}{clllllcll}
\hline
Target&\multicolumn{1}{c}{$\epsilon_2$}&\multicolumn{1}{c}{S$_{n}$}
&\multicolumn{1}{c}{DC}&\multicolumn{1}{c}{CN}&\multicolumn{1}{c}{DC+CN}
&\multicolumn{1}{c}{\%}&\multicolumn{1}{c}{$T_{1/2}$(n)}
&\multicolumn{1}{c}{$T_{1/2}$($\beta$)}\\
 & &\multicolumn{1}{c}{[MeV]}&\multicolumn{1}{c}{[mb]}
&\multicolumn{1}{c}{[mb]}&\multicolumn{1}{c}{[mb]}
& &\multicolumn{1}{c}{[s]}&\multicolumn{1}{c}{[s]}\\
\hline
$^{40}$S&+0.24&3.8238$^{\ddagger}$&0.4246&0.0851&0.5097&83&0.218&
8.8$\pm$2.2$^{\dagger}$\\
$^{42}$S&+0.30&3.3114$^{\ddagger}$&0.9466&0.0202&0.9668&98
&0.115&0.579$\pm$0.046$^{\dagger}$\\
$^{44}$S&+0.30&1.3344&0.014&0.0044&0.0184&76
&6.040&0.123$\pm$0.010$^{\dagger}$\\
$^{46}$Ar&$-$0.18&4.2590$^{\ddagger}$&0.5295&0.1203&0.6498&81&0.171&
8.4$\pm$0.6$^{\dagger}$\\
$^{48}$Ar&$-$0.22&1.7074&0.0427&0.0144&0.0571&75&1.946&0.11\\
$^{50}$Ar&$-$0.28&1.0804&0.0016&0.0030&0.0046&35&24.15&0.05\\
$^{56}$Ti&+0.13&2.1936$^{\ddagger}$&0.0147&0.1305&0.1452&10&0.038&
0.15$\pm$0.03$^{\dagger}$\\
$^{58}$Ti&$-$0.10&2.4244&0.0185&0.0797&0.0982&19&0.057&
0.047$\pm$0.010$^{\dagger}$\\
$^{60}$Ti&$-$0.02&2.1194&0.0165&0.0403&0.0568&29&0.098&0.054\\
$^{62}$Ti&$-$0.04&0.5878&0.0068&0.0030&0.0098&69&0.569&0.018\\
$^{64}$Ti&+0.06&0.4094&0.0013&0.0002&0.0015&87&3.561&0.039\\
$^{66}$Ti&+0.14&0.3914&0.0009&0.0002&0.0011&82&4.960&0.013\\
$^{62}$Cr&+0.30&2.9134&0.0119&0.3846&0.3965&$<$1&0.014&
0.187$\pm$0.015$^{\dagger}$\\
$^{64}$Cr&+0.05&1.8614&0.0170&0.0353&0.0523&33&0.106&
0.044$\pm$0.012$^{\dagger}$\\
$^{66}$Cr&+0.10&2.2374&0.0126&0.0673&0.0799&16&0.071&0.071\\
$^{68}$Cr&+0.16&1.8274&0.0129&0.0268&0.0397&32&0.140&0.026\\
$^{72}$Fe&+0.14&2.2474&0.0101&0.0572&0.0673&15&0.083&0.089\\
$^{74}$Fe&+0.07&2.0564&0.0104&0.0360&0.0464&22&0.120&0.052\\
$^{76}$Fe&+0.06&0.0584&0.0050&$3\cdot10^{-6}$&0.0050&$>$99&1.12&0.045\\
\hline
\end{tabular}
\end{center}
$^{\dagger}$ experimental $\beta$-decay half-life 
(Audi and Wapstra 1993, Pfeiffer et al. 2000)\\
$^{\ddagger}$ experimental $S_{\rm n}$ value
(Audi and Wapstra 1993)
\end{table}

From Table~II one can see nicely the importance of direct capture when 
approaching the magic neutron number $N=28$ (S, Ar), but also the 
increasing contribution of DC to the total cross section when approaching 
the neutron drip-line. 
As already mentioned above, {\it ''turning-points''} are reached in our 
r-process scenario for those isotopes where the half-life for capturing 
the next neutron becomes longer than the $\beta$-decay lifetime. Consequently, 
the nuclear-physics properties of $N=28$ $^{44}$S, $N=30$ $^{48}$Ar, $N=36$ 
$^{58}$Ti, $N=40$ $^{64}$Cr and $N=48$ $^{74}$Fe are of primary importance.

\section{Astrophysical Network Calculations}

In the late phase of massive stars, prior to the occurrence of rapid 
neutron-capture nucleosynthesis, reactions with charged particles such as 
(p,$\gamma$), (p,n), ($\alpha$,n) and ($\alpha$,$\gamma$), as well as their 
inverse reactions, ${\beta}^+$-decays and electron captures take place at 
high temperatures of $T_9\ge3$. The process paths as well as the 
development of the stars can be studied in full network calculations 
nowadays. The reliability of such calculations depends, among other things, 
also on the sophistication and internal consistency of the nuclear-physics 
input of the involved isotopes. Hence, in our nucleosynthesis calculations 
all nuclear input data, such as masses, neutron-separation energies, 
level-energies, their spin and parity, the ground-state shape, 
neutron-capture cross sections and decay-properties were taken into 
account. If available, experimental data were used, otherwise the input 
data were obtained from large-scale calculations using modern 
macroscopic-microsocpic models. 

As already pointed out in the introduction, the essential goal of our 
present investigation was to find a realistic nucleosynthesis 
process which reproduces simultaneously the complete Ca--Ti--Cr isotopic 
FUN anomalies observed in the EK-1-4-1 inclusion. The required stellar 
parameters, in particular neutron density (${\rho}_{\rm n}$), radiation-entropy 
($S \propto \,\, {T^3_9}/{\rho}$) and time scale of the process 
($\tau$), are then expected to give tight constraints on the possible 
astrophysical conditions within the chosen scenario. As in Freiburghaus 
et al. (1999), our calculations follow a hot blob of matter with entropy 
S: (1) Initially, it consists of neutrons, protons and some 
$\alpha$-particles in a ratio given by nuclear statistical equilibrium 
(NSE) for a specific proton-to-nucleon ratio $Y_{\rm e}$, and expands 
adiabatically and cools from $T_9\simeq 9$ to freeze-out conditions of 
$T_9 \simeq1$. (2) Thereby, the nucleons and $\alpha$-particles combine to 
heavier nuclei (typically Fe group), with some free neutrons and 
$\alpha$-particles remaining. (3) For high entropies, an $\alpha$-rich 
freeze-out from charged-particle reactions occurs for declining temperatures, 
producing nuclei in the mass range $A\simeq 80$--100. (4) Finally, below 
$T_9\simeq 3$ these nuclei can capture the remaining neutrons and undergo 
r-process-like processes. We consider a simple model of an adiabatically 
expanding homogeneous mass zone with an expansion timescale relevant to the 
supernova problem. It is generally accepted now that this approximation is 
justified by hydrodynamic and semi-analytical simulations of the evolution 
of the high-entropy wind of a SN. The calculations were performed for a grid 
of entropies ($S=3$, 10, 20,\dots, 300 $k_{\rm B}$/baryon) and 
proton-to-nucleon ratios ($Y_{\rm e}=0.42$, 0.45, 0.48), the latter 
''measuring'' the neutron-richness of the initial composition.    

Because of the chronological order of the so-called {\it $\alpha$-process}
and a neutron-capture process in the late phase of a massive star, our 
computer-code was organized in two steps. That part of the network which 
calculates the charged-particle reactions, was developed by F.-K.~Thielemann 
(1990) and extends from $_1$H to $_{46}$Pd. On the neutron-rich side all 
nuclei up to the neutron-drip line were included. On the proton-rich side, 
only the first three isotopes away from ${\beta}$-stability were taken into 
account. After the initial calculations of the $\alpha$-process, subsequent 
calculations were performed with the r-process code from C.~Freiburghaus 
for those cases where one or more free neutrons per seed-nucleus 
($Y_{\rm n}/Y_{\rm seed} \ge 1$) were still existing. A detailed description of 
the physical processes and the necessary system of equations is given in 
(Freiburghaus 1999). These r-process calculations are fully dynamical and 
are not restricted to a (n,$\gamma$)-($\gamma$,n) equilibrium, as were 
former calculations. 

\begin{figure}
\centerline{\psfig{file=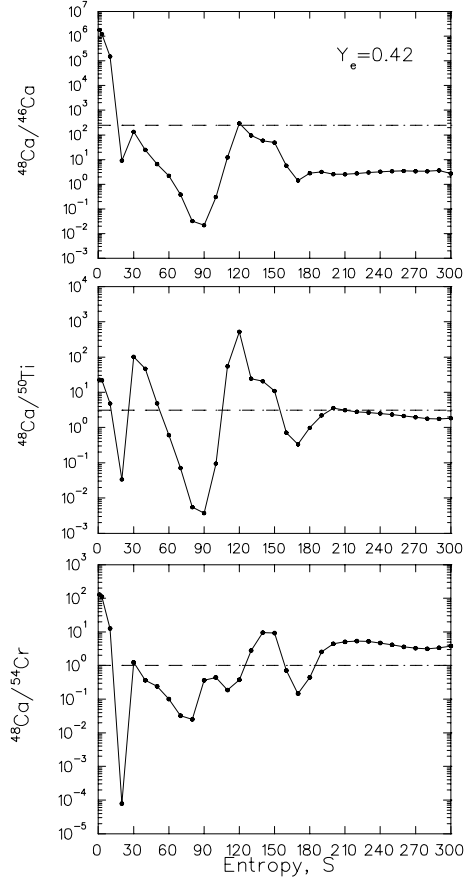,width=6.0cm}}
\caption[h]{Calculated isotope abundance ratios of $^{48}$Ca/$^{46}$Ca, 
$^{48}$Ca/$^{50}$Ti and $^{48}$Ca/$^{54}$Cr as a function of entropy. Here,  
a proton-to-nucleon ratio of $Y_{\rm e}=0.42$ has been chosen. The dashed lines 
represent the respective abundance ratios observed in the EK-1-4-1 inclusion 
of the Allende meteorite. }
\end{figure}  

\subsection{Isotope Abundances as a Function of Entropy}

In Fig.~1 the calculated isotope abundance ratios for $^{48}$Ca/$^{46}$Ca, 
$^{48}$Ca/$^{50}$Ti and $^{48}$Ca/$^{54}$Cr at $Y_{\rm e}=0.42$ are displayed 
as a function of entropy ($3\le S \le 300$). The dashed lines represent the 
values observed in EK-1-4-1; i.e. $^{48}$Ca/$^{46}{\rm Ca}\simeq 250$, 
$^{48}$Ca/$^{50}{\rm Ti} \simeq 3$, and $^{48}$Ca/$^{54}{\rm Cr} \simeq 1$. 
The isotopes $^{48}$Ca, $^{50}$Ti and $^{54}$Cr are the most neutron-rich 
stable ones for each element, and all three are significantly overabundant 
compared to ''normal'' solar matter. \hfill  Moreover, even their \hfill 
ratios among each 
\hfill other are \hfill enriched in \hfill $^{48}$Ca by

\begin{center}
\begin{minipage}{6cm}
\centerline{\psfig{file=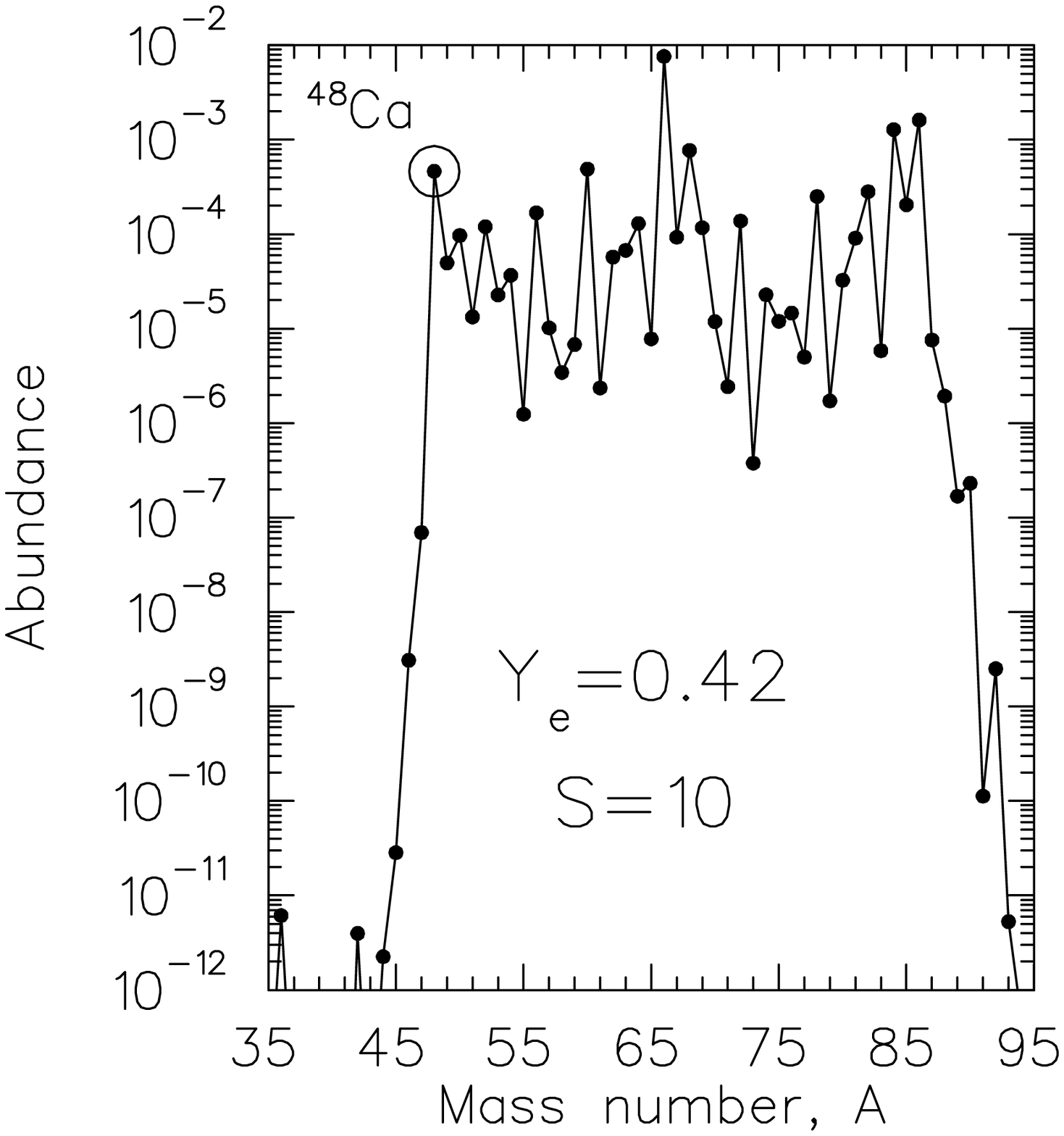,width=5.75cm}}
\end{minipage}
\hfill
\begin{minipage}{6cm}
\centerline{\psfig{file=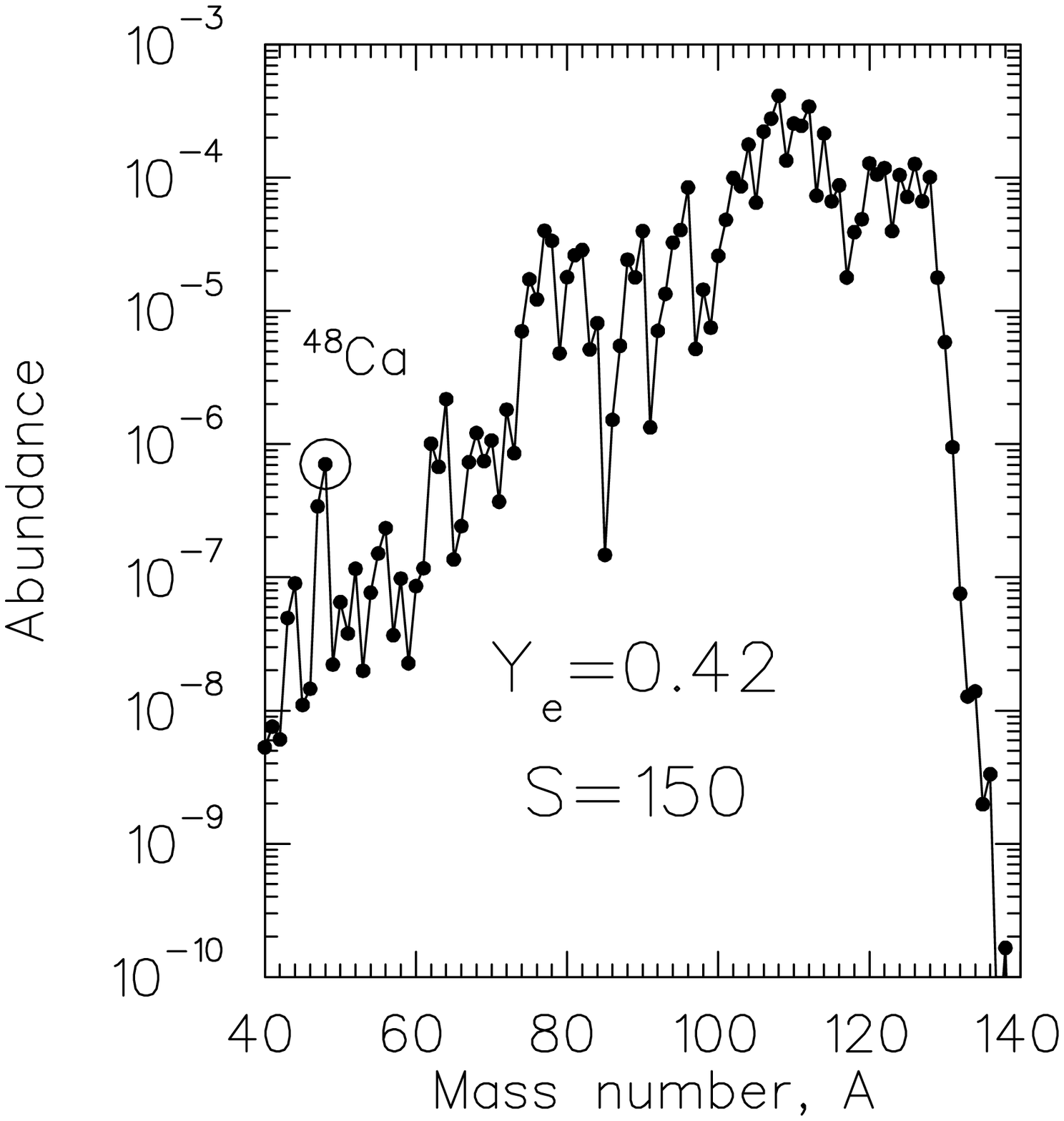,width=5.75cm}}
\end{minipage}
\vspace{5mm}
\begin{minipage}{130mm}
Fig.~2. Isotopic abundances calculated for a proton-to-nucleon ratio 
$Y_{\rm e}=0.42$ and entropies of $S = 10$ $k_{\rm B}$/baryon (left part) 
and $S = 150$ $k_{\rm B}$/baryon (right part).
\end{minipage}
\end{center}

\noindent
factors of 3--4 compared to solar. 
Therefore, over and above a reproduction of the abundances of the stable 
isotopes within each Z-chain, the most sensitive test of our entropy-based 
model will be the above abundance ratios.

As can be seen from Figure~1, there are two entropy regions for which these 
isotope abundance ratios in EK-1-4-1 can be reproduced, one at low entropies 
($S\simeq 10$ $k_{\rm B}$/baryon) and the other at high entropies ($S \simeq 
150$ $k_{\rm B}$/baryon). In the first case, the Ca--Ti--Cr 
isotopes are synthesized in an $\alpha$-rich freeze-out where essentially no 
free neutrons are available for subsequent neutron captures. Hence, the 
neutron-rich stable isotope $^{48}$Ca, for example, is formed {\it directly}. 
This is in agreement with the earlier study of Meyer, Krishnan and Clayton
(Meyer 1996). In the second case, already some amount of free neutrons exists 
that can start a kind of weak r-process after the charged-particle freeze-out. 
Under such conditions, $^{48}$Ca is mainly generated by $\beta$-decay of its 
progenitor $^{48}$Ar. And $^{44}$S, indeed, acts as a {\it ''turing-point''} 
nucleus, thus avoiding the production of large amounts of $^{46}$S the main 
progenitor of stable isobaric $^{46}$Ca.

It may be of further interest to compare the whole abundance distributions 
produced under the above two entropy conditions (see Fig.~2). In 
principal agreement with the results obtained by Freiburghaus et al. (1999), 
for low entropies sizeable abundances from the freeze-out of an $\alpha$-rich 
Si-QSE (quasi nuclear statistical equilibrium) are obtained only in a limited 
mass range; i.e. within $45\le A \le 95$. For higher entropies, abundances 
from the freeze-out of a neutron-rich Fe-QSE extend up to heavier masses. 
They show an increasing pattern from $A \simeq 40$ to $A \simeq120$, with a 
sudden drop around $A \simeq 130$ where the neutron-capture path reaches 
$N=82$ neutron-magic isotopes. There are also considerable differences in 
the absolute abundances of the two scenarios, shown in Fig.~2 in the units 
of the solar Si abundance (${\rm Si}=10^6$). At low entropies, the even-even 
Ca--Ti--Cr isotopes of interest are produced at a level of about 10$^{-4}$, 
whereas for the higher entropies they lie roughly three orders of magnitude 
lower. 

It may also be of interest to consider abundance ratios of e.g. $^{48}$Ca 
with heavier neutron-rich stable isotopes expected -- and, indeed, partly 
observed in CaAl-rich inclusions -- to be overabundant compared to solar, 
such as $^{58}$Fe, $^{64}$Ni, $^{66}$Zn and $^{96}$Zr. As may already be 
inferred from a fleeting glance of Fig.~2, there are large differences for 
the low- and high-entropy scenarios. If some tiny amount of  EK-1-4-1 
material could still be found, those different abundance patterns might 
motivate further experimental studies with todays much more sensitive 
techniques. To give an example, in case of the formation of the Ca--Ti--Cr 
abundances in a Si-QSE (see, S=10) the abundance ratios of 
Y($^{90-96}$Zr)/Y($^{48}$Ca) 
should lie between 5$\cdot$10$^{-4}$ and 5$\cdot$10$^{-10}$, i.e. 
{\it negligible}, whereas for an Fe-QSE followed by a weak r-process (S=150) 
the above ratios would be orders of magnitude larger (in detail, 56, 1.7, 11, 
45, 120 for $^{90,91,92,94,96}$Zr, respectively). Under these latter 
conditions, the calculated Y($^{91,92}$Zr)/Y($^{90}$Zr) isotopic ratios 
would be exactly the solar r-abundance (Y$_{r,\odot}$) ratios (K\"appeler 
et al. 1989, Arlandini et al. 1999), whereas the 
$^{94,96}$Zr/$^{90}$Zr ratios already show considerable overabundances 
of the heaviest Zr isotopes compared to Y$_{r,\odot}$ by factors of about 8 
and 12, respectively. 

In the context of the question, which of our two nucleosynthesis scenarios 
-- the low- or the high-entropy approach -- has most likely produced the FUN 
anomalies in EK-1-4-1, it is interesting to remember that in fact the very 
first measurements were made for Ba, Nd and Sm (Lugmair et al. 1978, 
McCulloch and Wasserburg 1978a, 1978b). They found excesses of the unshielded 
heaviest isotopes, which are generally believed to be pure r-process products. 
Within our entropy-based parameter study, only the high-entropy scenario is 
able to produce isotopes beyond A$\simeq$130 with measurable abundances 
(Y(Z$\ge$56)$\ge$10$^{-10}$). At the example of the Sm isotopic ratios, 
relative to r-only $^{149}$Sm, in Table III we compare the weighted averages 
of the experimental values obtained by Lugmair et al. (1978) and McCulloch 
and Wasserburg (1978) for EK-1-4-1 with those of the solar-system s- and 
r-process abundances and our predictions for different entropies in the range 
S=150--230. It is immediately evident that the measured EK-1-4-1 values are 
undoubtedly {\bf} not of s-process origin; however, they are also not fully 
compatible with the entropy- (neutron-density) mix required to explain the 
Y$_{r,\odot}$ pattern. Our calculations show that an r-process-like scenario 
with a {\it limited range} of entropies is able to reproduce the observed 
values. When starting with S=150, the absolute abundance of $^{149}$Sm is of 
the order of Y($^{149}$Sm)$\simeq$2$\cdot$10$^{-13}$ and seems to be still 
overproduced, whereas the two heaviest Sm isotopes are still underproduced 
relative to EK-1-4-1. This picture changes gradually with increasing 
entropies. The absolute abundances of all Sm nuclides increase up to 
S$\simeq$210 (e.g., Y($^{149}$Sm)$\simeq$4$\cdot$10$^{-10}$ for S=170, 
3$\cdot$10$^{-7}$ for S=190, 1$\cdot$10$^{-5}$ for S=210) and then decrease 
again. In parallel, with increasing entropy more and more r-material is 
shifted to the heavier Sm isotopes. This is clearly reflected by the abundance 
ratios Y($^{152,154}$Sm)/Y($^{149}$Sm) in the table. Not only when considering 
the ratios, but also when using the {\it absolute} Sm abundances {\it best} 
agreement with EK-1-4-1 is obtained around S$\simeq$190--200. These are 
conditions just at the upper end of the high-entropy region where also the 
correlated Ca--Ti--Cr anomalies in this inclusion could be reproduced. 
It is worth to be mentioned in this context, that a quite similar picture is 
observed for the Nd isotopic abundance ratios measured by Lugmair et al. 
(1978) and McCulloch and Wasserburg (1978 a). A more detailed exploration of 
the astrophysical conditions to produce those isotopes in EK-1-4-1 and other 
meteoritic inclusions is presently underway and will be published elsewhere.
 
\begin{table}             
\caption{Comparison of Sm isotopic abundance ratios observed in EK-1-4-1
(Lugmair et al. 1978, McCulloch and Wasserburg 1978b) with solar-system
s- and r-process abundances, and with our network calculations for different
entropies.
}
\begin{center}
\begin{tabular}{ccccccccc}
\hline
Isotope & Y$_{s,\odot}$ & Y$_{r,\odot}$ & Y$_{EK-1-4-1}$ & 
\multicolumn{5}{c}{Network calculations} \\
abund.-ratio &$^{a)}$ &$^{b)}$ &$^{c)}$ & S=150 & S=170 & S=190 & S=210 & S=230\\
\hline
$^{147}$Sm/$^{149}$Sm & 1.92 & 1.01 & 0.99$\pm$0.10 & 3.05 & 1.82 & 1.23 &
0.93 & 0.87 \\
$^{152}$Sm/$^{149}$Sm & 3.50 & 1.71 & 0.74$\pm$0.10 & 0.25 & 0.22 & 0.77 &
2.18 & 2.18 \\
$^{154}$Sm/$^{149}$Sm & 0.088 & 1.87 & 1.00$\pm$0.11 & 0.035 & 0.16 & 0.47 &
1.54 & 1.81 \\
\hline
\end{tabular}
\end{center}
$^{a)}$ K\"appeler et al. 1990\\
$^{b)}$ Arlandini et al. 1999\\
$^{c)}$ Lugmair et al. 1978, McCulloch and Wasserburg 1978b
\end{table}

\subsection{Possible Astrophysical Scenarios} 

We have shown that resultant abundances of neutron-rich stable isotopes in the 
$^{48}$Ca to $^{66}$Zn region can be produced in matter expanding from high 
temperature and density for two different entropy regions. In agreement with 
Meyer et al. (Meyer 1996), we find that for low entropies 
($S \simeq 10$ $k_{\rm B}$/baryon) these isotopes are synthesized in the 
$\alpha$-rich freeze-out from a Si-QSE. In this case, the key nucleus $^{48}$Ca 
is synthesized {\it directly} without significant neutron captures. Such 
conditions are found in Type Ia supernovae (SN~Ia), which are certainly 
responsible for the production of the ''bulk'' $^{48}$Ca material in our solar 
system, as outlined by Meyer, Krishnan and Clayton.      
 
We have furthermore shown that the Ca--Ti--Cr anomalies in EK-1-4-1 can also 
be reproduced with higher entropies of S$\simeq$150--190 $k_{\rm B}$/baryon, 
however with considerably lower yields than with low entropies. Under these 
conditions, neutron-rich matter is formed by the freeze-out of an 
$\alpha$-rich Fe-QSE with subsequent r-process-like neutron captures. In 
this case, the abundance distribution is much larger and ranges up to the 
rare-earth region. A realistic astrophysical site for such a scenario would 
be the high-entropy bubble at the outer core of a type II supernova (SN~II). 
Meyer et al. (1996) conclude from their detailed NSE network calculations 
that $^{48}$Ca does not survive such an environment, because the material 
freezes out with too few nuclei at low masses. In their model, the QSE nearly 
exclusively produces nuclei heavier than $^{48}$Ca; therefore, they make the 
rather stong statement: {\it ''We therefore rule out this Type II site as an 
origin for $^{48}$Ca, even in matter having the neutron richness of 
$^{48}$Ca.''}. Here, we disagree with Meyer et al. (1996). In our network 
calculations, maybe due to the considerably improved nuclear-physics data 
input for neutron-rich isotopes, we still obtain measurable quantities of 
nuclides in the $^{48}$Ca region {\it simultaneously} with overabundances 
of the heaviest stable isotopes up to Nd and Sm, also observed in EK-1-4-1
(Lugmair et al. 1978, McCulloch and Wasserburg 1978a,1978b).
Typical yields for low-mass isotopes with S=150 $k_{\rm B}$/baryon are  
$1 \cdot 10^{-10} M_{\odot}$ for $^{46}$Ca,  
$4 \cdot 10^{-8} M_{\odot}$ for $^{48}$Ca,  
$2 \cdot 10^{-9} M_{\odot}$ for $^{50}$Ti and  
$3 \cdot 10^{-8} M_{\odot}$ for $^{54}$Cr. This is 
about 2 to 3 orders of magnitude less than in SN~Ia.  Moreover, SN~II events 
occur approximately ten times less than SN~Ia. Taken together, the 
contribution of synthesized and ejected matter from SN~II is a factor 10$^3$ 
to 10$^4$ less than the ''bulk'' material from SN~Ia.  Hence, we absolutely 
agree with Meyer et al. (1996), that SN~II do not produce enough low-mass 
material to generate the solar amounts of $^{48}$Ca, $^{50}$Ti and  $^{54}$Cr. 
However, the amount of matter ejected by Type II supernovae may well be 
sufficient to explain the observations in the specific SN condensate 
EK-1-4-1. In an attempt to explain the correlation of overabundances of 
neutron-rich isotopes over a wide range of elements in CaAl-rich inclusions, 
Meyer et al. (1996) argue that this material, which was formed in the solar 
accretion disk, contains {\it ''components not only from many Type Ia 
supernovae but from Type II supernovae as well.''}. In our high-entropy 
approach, however, {\bf all} FUN anomalies, from $^{48}$Ca up to $^{154}$Sm 
observed in EK-1-4-1, are produced simultaneously in the high-entropy bubble 
scenario of a SN~II. Therefore, extending the early notation of Wasserburg
(Lee 1978, Niederer 1980), we would now speak not only about correlated 
Ca--Ti  anomalies, but about correlated overabundances of neutron-rich 
Ca--Ti--Cr--Fe--Ni--Zn$\dots$Zr$\dots$Ba--Nd--Sm isotopes in EK-1-4-1. 

In full agreement with Meyer, Krishnan and Clayton we conclude that certainly 
much work remains to be done, although we have also learned a great 
deal. It will therefore be interesting to see how our insights into FUN 
syntheses will fit in with future astrophysical models.

\acknowledgements
Financial support from the German DFG is gratefully acknowledged. The authors 
thank U.~Ott for helpful comments.

\end{document}